\documentclass[runningheads]{llncs}
\usepackage[T1]{fontenc}
\usepackage{graphicx}
\usepackage{multirow}
\usepackage{marvosym}
\usepackage{cite}
\usepackage[colorlinks,linkcolor=blue]{hyperref}
\begin{document}
\title{MUVF-YOLOX: A Multi-modal Ultrasound Video Fusion Network for Renal Tumor Diagnosis}
\titlerunning{A MUVF Network for Renal Tumor Diagnosis}
%
\author{Junyu Li\inst{1,2,3} \and
Han Huang\inst{1,2,3} \and
Dong Ni\inst{1,2,3} \and
Wufeng Xue\inst{1,2,3} \and
Dongmei Zhu\inst{4}\textsuperscript{(\Letter)} \and 
Jun Cheng\inst{1,2,3}\textsuperscript{(\Letter)}}
%
%
\institute{National-Regional Key Technology Engineering Laboratory for Medical Ultrasound, School of Biomedical Engineering, Shenzhen University Medical School, Shenzhen University, Shenzhen, China \\ \email{chengjun583@qq.com} \and
Medical UltraSound Image Computing (MUSIC) Lab, Shenzhen University, Shenzhen, China \and
Marshall Laboratory of Biomedical Engineering, Shenzhen University, Shenzhen, China \and
Department of Ultrasound, The Affiliated Nanchong Central Hospital of North Sichuan Medical College, Nanchong, China \\ 
\email{zdm596987@gmail.com }\\
}
%
\maketitle              
\begin{abstract}
Early diagnosis of renal cancer can greatly improve the survival rate of patients. Contrast-enhanced ultrasound (CEUS) is a cost-effective and non-invasive imaging technique and has become more and more frequently used for renal tumor diagnosis. 
However, the classification of benign and malignant renal tumors can still be very challenging due to the highly heterogeneous appearance of cancer and imaging artifacts.
Our aim is to detect and classify renal tumors by integrating B-mode and CEUS-mode ultrasound videos.
To this end, we propose a novel multi-modal ultrasound video fusion network that can effectively perform multi-modal feature fusion and video classification for renal tumor diagnosis.
The attention-based multi-modal fusion module uses cross-attention and self-attention to extract modality-invariant features and modality-specific features in parallel.
In addition, we design an object-level temporal aggregation (OTA) module that can automatically filter low-quality features and efficiently integrate temporal information from multiple frames to improve the accuracy of tumor diagnosis.
Experimental results on a multicenter dataset show that the proposed framework outperforms the single-modal models and the competing methods.
Furthermore, our OTA module achieves higher classification accuracy than the frame-level predictions. 
Our code is available at \url{https://github.com/JeunyuLi/MUAF}. 
\keywords{Multi-modal Fusion \and Ultrasound Video \and Object Detection \and Renal Tumor.}
\end{abstract}
\section{Introduction}
\par
Renal cancer is the most lethal malignant tumor of the urinary system, and the incidence is steadily rising \cite{ljungberg2019european}.
Conventional B-mode ultrasound (US) is a good screening tool but can be limited in its ability to characterize complicated renal lesions. Contrast-enhanced ultrasound (CEUS) can provide information on microcirculatory perfusion.
Compared with CT and MRI, CEUS is radiation-free, cost-effective, and safe in patients with renal dysfunction. Due to these benefits, CEUS is becoming increasingly popular in diagnosing renal lesions. 
However, recognizing important diagnostic features from CEUS videos to diagnose lesions as benign or malignant is non-trivial and requires lots of experience. 

To improve diagnostic efficiency and accuracy, many computational methods were proposed to analyze renal US images and could assist radiologists in making clinical decisions \cite{george2022analysis}. However, most of these methods only focused on conventional B-mode images. 
In recent years, there has been increasing interest in multi-modal medical image fusion\cite{azam2022review}.
Directly concatenation and addition were the most common methods, such as \cite{liu2017medical, fang2022weighted, chen2019multi}.
These simple operations might not highlight essential information from different modalities. 
Weight-based fusion methods generally used an importance prediction module to learn the weight of each modality and then performed sum, replacement, or exchange based on the weights \cite{huang2022personalized, yang2019comprehensive, wang2020deep, wang2022channel}. 
Although effective, these methods did not allow direct interaction between multi-modal information. 
To address this, attention-based methods were proposed. They utilized cross-attention to establish the feature correlation of different modalities and self-attention to focus on global feature modeling \cite{li2022transiam, xu2022remixformer}.
Nevertheless, we prove in our experiments that these attention-based methods may have the potential risks of entangling features of different modalities.

In practice, experienced radiologists usually utilize dynamic information on tumors’ blood supply in CEUS videos to make diagnoses\cite{kapetas2019quantitative}. Previous researches have proved that temporal information is effective in improving the performance of deep learning models. Lin et al.\cite{lin2022new} proposed a network for breast lesion detection in US videos by aggregating temporal features, which outperformed other image-based methods. Chen et al.\cite{chen2021domain} showed that CEUS videos can provide more detailed blood supply information of tumors allowing a more accurate breast lesion diagnosis than static US images. 

In this work, we propose a novel multi-modal US video fusion network (MUVF-YOLOX) based on CEUS videos for renal tumor diagnosis. Our main contributions are fourfold. 
(1) To the best of our knowledge, this is the first deep learning-based multi-modal framework that integrates both B-mode and CEUS-mode information for renal tumor diagnosis using US videos. 
(2) We propose an attention-based multi-modal fusion (AMF) module consisting of cross-attention and self-attention blocks to capture modality-invariant and modality-specific features in parallel. 
(3) We design an object-level temporal aggregation (OTA) module to make video-based diagnostic decisions based on the information from multi-frames. 
(4) We build the first multi-modal US video datatset containing B-mode and CEUS-mode videos for renal tumor diagnosis. 
Experimental results show that the proposed framework outperforms single-modal, single-frame, and other state-of-the-art methods in renal tumor diagnosis.

\section{Methods}
\subsection{Overview of Framework}
\begin{figure}
    \centering
    \includegraphics[width=1\textwidth]{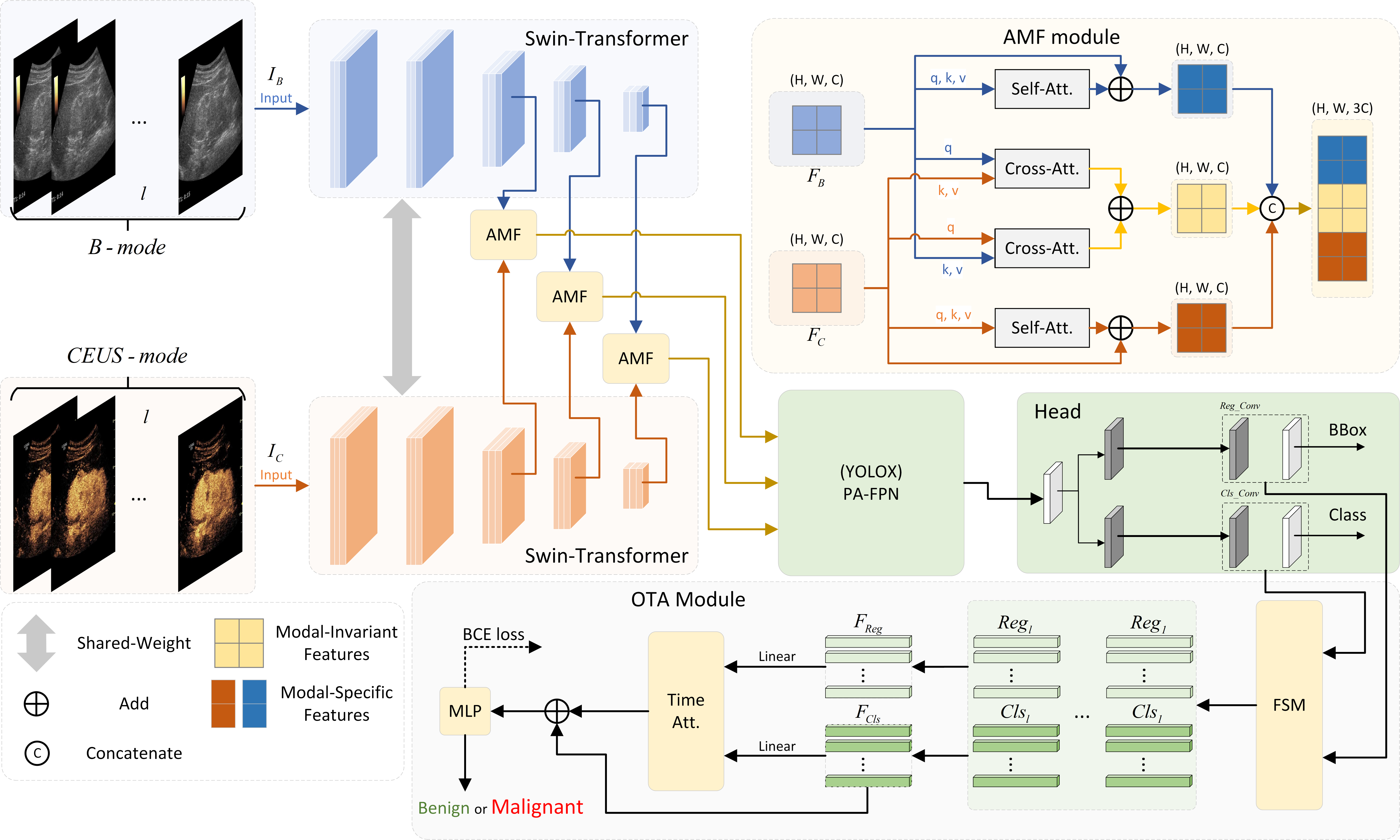}
    \caption{Framework of MUVF-YOLOX. AMF module is used to fuse multi-modal features. OTA module is used to classify the tumor as benign or malignant based on videos. FSM means feature selection module.}
    \label{Fig: network}
\end{figure}

The proposed MUVF-YOLOX framework is shown in Fig.~\ref{Fig: network}. It can be divided into two stages: 
single-frame detection stage and video-based diagnosis stage.
(1) In the single-frame detection stage, the network predicts the tumor bounding box and category on each frame in the multi-modal CEUS video clips. 
Dual-branch backbone is adopted to extract the features from two modalities and followed by the AMF module to fuse these features. 
During the diagnostic process, experienced radiologists usually take the global features of US images into consideration\cite{zhu2022contrast}.
Therefore, we modify the backbone of YOLOX from CSP-Darknet to Swin-Transformer-Tiny, which is a more suitable choice by the virtue of its global modeling capabilities\cite{wang2021transbts}.
(2) In the video-based diagnosis stage, the network automatically chooses high-confidence region features of each frame according to the single-frame detection results and performs temporal aggregation to output a more accurate diagnosis. 
The above two stages are trained successively. 
We first perform a strong data augmentation to train the network for tumor detection and classification on individual frames. 
After that, the first stage model is switched to the evaluation mode and predicts the label of each frame in the video clip. 
Finally, we train the OTA module to aggregate the temporal information for precise diagnosis. 
\subsection{Dual-Attention Strategy for Multimodal Fusion}  
Using complementary information between multi-modal data can greatly improve the precision of detection. Therefore, we propose a novel AMF module to fuse the features of different modalities. As shown in Fig.~\ref{Fig: network}, the features of each modality will be input into cross-attention and self-attention blocks in parallel to extract modality-invariant features and modality-specific features simultaneously. 

Taking the B-mode as an example, we first map the B-mode features $F_{B}$ and the CEUS-mode features $F_{C}$ into $(Q_B, K_B, V_B)$ and $(Q_C, K_C, V_C)$ using linear projection. 
Then cross-attention uses scaled dot-product to calculate the similarity between $Q_B$ and $K_C$.
The similarity is used to weight $V_C$.
Cross-attention extracts modality-invariant features through correlation calculation but ignores modality-specific features in individual modalities. 
Therefore, we apply self-attention in parallel to highlight these features. 
The self-attention calculates the similarity between $Q_B$ and $K_B$ and then uses the similarity to weight $V_B$. 
Similarly, the features of the CEUS modality go through the same process in parallel. 
Finally, we merge the two cross-attention outputs by addition since they are both invariant features of two modalities and concatenate the obtained sum and the outputs of the two self-attention blocks. 
The process mentioned above can be formulated as follows:
\begin{equation}
F_{invar} = Softmax(\frac{{Q_B}{K^T_C}}{\sqrt{d}})V_C + Softmax(\frac{{Q_C}{K^T_B}}{\sqrt{d}})V_B
\end{equation}
\begin{equation}
F_{B-spec} = Softmax(\frac{{Q_B}{K^T_B}}{\sqrt{d}})V_B + F_{B}
\end{equation}
\begin{equation}
F_{C-spec} = Softmax(\frac{{Q_C}{K^T_C}}{\sqrt{d}})V_C + F_{C}
\end{equation}
\begin{equation}
F_{AMF} = Concat(F_{B-spec}, F_{invar}, F_{C-spec})
\end{equation}
where, $F_{invar}$ represents the modality-invariant features. $F_{B-spec}$ and $F_{C-spec}$ represent the modal-specific features of B-mode and CEUS-mode respectively. $F_{AMF}$ is the output of the AMF module.

\subsection{Video-level Decision Generation} 
In clinical practice, the dynamic changes in US videos provide useful information for radiologists to make diagnoses. 
Therefore, we design an OTA module that aggregates single-frame renal tumor detection results in temporal dimension for diagnosing tumors as benign and malignant.
First, we utilize a feature selection module\cite{shi2022yolov} to select high-quality features of each frame from the $Cls\_conv$ and $Reg\_conv$ layers.
Specifically, we select the top 750 grid cells on the prediction grid according to the confidence score.
Then, 30 of the top 750 grid cells are chosen by the non-maximum suppression algorithm for reducing redundancy. 
The features are finally picked out from the $Cls\_conv$ and $Reg\_conv$ layers guided by the positions of the top 30 grid cells.
Let $F_{Cls}=\{Cls_1, Cls_2,...Cls_l\}$ and $F_{Reg}=\{Reg_1, Reg_2,...Reg_l\}$ denote the above obtained high-quality features from $l$ frames.
After feature selection, we aggregate the features in the temporal dimension by time attention. 
$F_{Cls}$ and $F_{Reg}$ are mapped into $(Q_{Cls}, K_{Cls}, V_{Cls})$ and $(Q_{Reg}, K_{Reg})$ via linear projection. 
Then, we utilize scaled dot-product to compute the attention weights of $V_{Cls}$ as:
\begin{equation}
Time\_Att. = [Softmax(\frac{{Q_{Cls}}{K^T_{Cls}}}{\sqrt{d}}) + Softmax(\frac{{Q_{Reg}}{K^T_{Reg}}}{\sqrt{d}})]V_{Cls}
\end{equation}
\begin{equation}
F_{temp} = Time\_Att. + F_{Cls}
\end{equation}
After temporal feature aggregation, $F_{temp}$ is fed into a multilayer perceptron head to predict the class of tumor.
\section{Experimental Results}
\subsection{Materials and Implementations}
\begin{figure}
    \centering
    \includegraphics[width=1\textwidth]{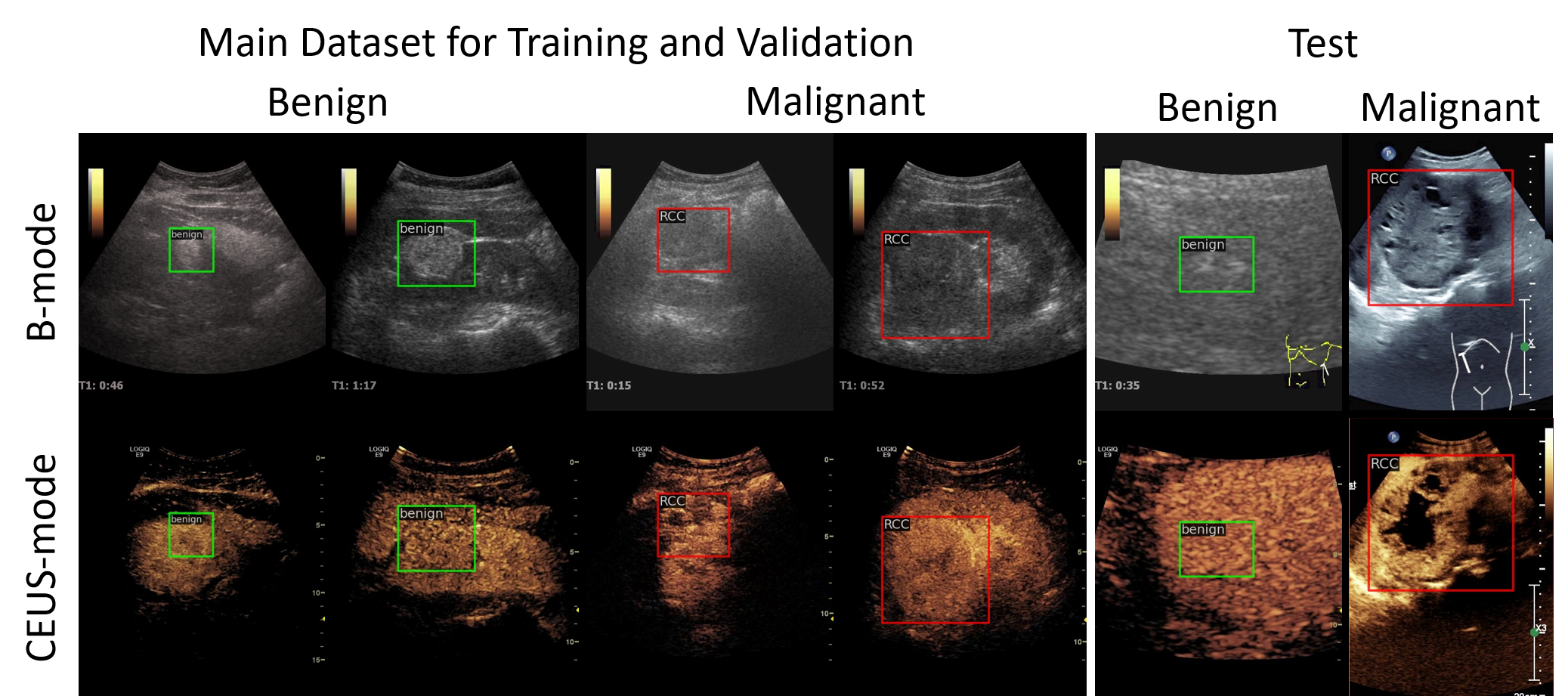}
    \caption{Examples of the annotated B-mode and CEUS-mode US images. }
    \label{Fig: Dataset}
\end{figure}

We collect a renal tumor US dataset of 179 cases from two medical centers, which is split into the training and validation sets. 
We further collect 36 cases from the two medical centers mentioned above (14 benign cases) and another center (center A, 22 malignant cases) to form the test set. 
Each case has a video with simultaneous imaging of B-mode and CEUS-mode. 
Some examples of the images are shown in Fig.~\ref{Fig: Dataset}. There is an obvious visual difference between the images from center A (last column in Fig.~\ref{Fig: Dataset}) and the other two centers, which raises the complexity of the task but can better verify our method's generalization ability.
More than two radiologists with ten years of experience manually annotate the tumor bounding box and class label at the frame level.
Each case has 40-50 labeled frames, and these frames cover the complete contrast-enhanced imaging cycle. 
The number of cases and annotated frames is summarized in Tab.~\ref{Data details}.

Weights pre-trained from ImageNet are used to initialize the Swin-Transformer backbone.
Data augmentation strategies are applied synchronously to B-mode and CEUS-mode images for all experiments, including random rotation, mosaic, mixup, and so on. 
All models are trained for 150 epochs. 
The batch size is set to 2. 
We use the SGD optimizer with a learning rate of 0.0025. The weight decay is set to 0.0005 and the momentum is set to 0.9. In the test phase, we use the weights of the best model in validation to make predictions. All Experiments are implemented in PyTorch with an NVIDIA RTX A6000 GPU.
$AP_{50}$ and $AP_{75}$ are used to assess the performance of single-frame detection. Accuracy and F1-score are used to evaluate the video-based tumor diagnosis. 
\begin{table}[]\centering
\caption{The details of our dataset. Number of cases in brackets.}\label{Data details}
\begin{tabular}{|c|c|c|c|}
\hline
Category  & Training     & Validation & Test        \\ \hline
Benign    & 2775(63)  & 841(16)    & 875(14)     \\ \cline{1-1}
Malignant & 4017(81)  & 894(19)    & 1701(22)    \\ \cline{1-1}
Total     & 6792(144) & 1735(35)   & 2576(36)    \\ \hline
\end{tabular}
\end{table}
\subsection{Ablation Study}
\subsubsection{Single-frame Detection.}We explore the impact of different backbones in YOLOX and different ways of multi-modal fusion. 
As shown in Tab.~\ref{ablation}, using Swin-Transformer as the backbone in YOLOX achieves better performance than the original backbone while reducing half of the parameters.
The improvement may stem from the fact that Swin-Transformer has a better ability to characterize global features, which is critical in US image diagnosis. 
In addition, we explore the role of cross-attention and self-attention blocks in multi-modal tasks, as well as the optimal strategy for combining their outputs.
Comparing row 5 with row 7 and row 8 in Tab.~\ref{ablation}, the dual-attention mechanism outperforms the single cross-attention. 
It indicates that we need to pay attention to both modality-invariant and modality-specific features in our multi-modal task through cross-attention and self-attention blocks. 
However, "CA+SA" (row 6 in Tab.~\ref{ablation}) obtains inferior performance than "CA" (row 5 in Tab.~\ref{ablation}). 
We conjecture that connecting the two attention modules in series leads to the entanglement of modality-specific and modality-invariant information, which would disrupt the model training.
On the contrary, the "CA//SA" method, combining two attention modules in parallel, enables the model to capture and digest modality-specific and modality-invariant features independently. 
For the same reason, we concatenate the outputs of the attention blocks rather than summing, which further avoids confusing modality-specific and modality-invariant information.
Therefore, the proposed method achieves the best performance.
\begin{table}[]\centering
\caption{The results of ablation study.
"CA" and "SA" denote cross-attention and self-attention respectively. "//" and "+" mean parallel connection and series connection.}\label{ablation}
\begin{tabular}{|c|c|cc|cc|cc|}
\hline
\multirow{2}{*}{Modal}                             & \multirow{2}{*}{Network}                                           & \multicolumn{2}{c|}{Validation}           & \multicolumn{2}{c|}{Test}                                              & \multicolumn{1}{c|}{\multirow{2}{*}{\begin{tabular}[c]{@{}c@{}}Flops\\ (GLOPS)\end{tabular}}} & \multirow{2}{*}{\begin{tabular}[c]{@{}c@{}}Params\\ (M)\end{tabular}} \\ \cline{3-6}
                                                   &                                                                    & \multicolumn{1}{c|}{$AP_{50}$} & $AP_{75}$          & \multicolumn{1}{c|}{$AP_{50}$}         & $AP_{75}$                               & \multicolumn{1}{c|}{}                                                                         &                                                                       \\ \hline
\multirow{2}{*}{B-mode}                            & YOLOX\cite{ge2021yolox}                                                              & \textbf{60.7}             & \textbf{39.7} & 48.5                              & 19.6                               & 140.76                                                                                        & 99.00                                                                 \\ \cline{2-2}
                                                   & Swin-YOLOX                                                         & 59.6                      & 38.0          & \textbf{58.1}                     & \textbf{22.1}                      & \textbf{61.28}                                                                                & \textbf{45.06}                                                        \\ \hline
\multirow{2}{*}{CEUS-mode}                         & YOLOX\cite{ge2021yolox}                                                              & 52.3                      & 29.7          & 49.1                              & \textbf{17.8}                               & 140.76                                                                                        & 99.00                                                                 \\ \cline{2-2}
                                                   & Swin-YOLOX                                                         & \textbf{60.1}             & \textbf{30.6} & \textbf{52.1}                     & 14.1                      & \textbf{61.28}                                                                                & \textbf{45.06}                                                        \\ \hline
\multicolumn{1}{|l|}{\multirow{4}{*}{Multi-modal}} & \begin{tabular}[c]{@{}c@{}}CA\\ (CMF\cite{xu2022remixformer})\end{tabular}                 & 81.4                      & 54.2          & 75.2                              & 35.2                               & 103.53                                                                                        & 51.26                                                                 \\ \cline{2-2}
\multicolumn{1}{|l|}{}                             & \begin{tabular}[c]{@{}c@{}}CA+SA\\ (TMM\cite{li2022transiam})\end{tabular}              & 80.8                      & 52.7          & 74.3                              & 37.0                               & 109.19                                                                                        & 57.46                                                                 \\ \cline{2-2}
\multicolumn{1}{|l|}{}                             & \begin{tabular}[c]{@{}c@{}}CA//SA\\ (Ours w/o Concat)\end{tabular} & 82.0                      & 56.9          & 74.6                              & 35.0                               & 109.19                                                                                        & 57.46                                                                 \\ \cline{2-8} 
\multicolumn{1}{|l|}{}                             & \textbf{Ours}                                                      & \textbf{82.8}             & \textbf{60.6} & \multicolumn{1}{l}{\textbf{79.5}} & \multicolumn{1}{l|}{\textbf{39.2}} & 117.69                                                                                        & 66.76                                                                 \\ \hline
\end{tabular}
\end{table}
\subsubsection{Video-based Diagnosis.}We investigate the performance of the OTA module for renal tumor diagnosis in multi-modal videos.
We generate a video clip with $l$ frames from annotated frames at a fixed interval forward.
As shown in Tab.~\ref{video}, gradually increasing the clip length can effectively improve the accuracy.
This suggests that the multi-frame model can provide a more comprehensive characterization of the tumor and thus achieves better performance.  
Meanwhile, increasing the sampling interval tends to decrease the performance (row 4 and row 5 in Tab.~\ref{video}). 
It indicates that continuous inter-frame information is beneficial for renal tumor diagnosis. 
\begin{table}[]\centering
\caption{The results of video-based diagnosis. }\label{video}
\begin{tabular}{|c|c|cc|cc|}
\hline
\multirow{2}{*}{Clip Length} & \multirow{2}{*}{Sampling Interval} & \multicolumn{2}{c|}{Validation}              & \multicolumn{2}{c|}{Test}                    \\ \cline{3-6} 
                          &                              & \multicolumn{1}{c|}{Accuracy(\%)} & F1-score(\%)  & \multicolumn{1}{c|}{Accuracy(\%)} & F1-score(\%)  \\ \hline
1                         & 1                            & 81.6                         & 81.6          & 90.3                         & 89.2          \\
2                         & 1                            & 82.1                         & 82.1          & 90.5                         & 89.3          \\
2                         & 2                            & 82.9                         & 82.9          & 90.0                         & 88.7          \\
4                         & 1                            & 83.7                         & 83.7          & \textbf{91.0}                & \textbf{90.0} \\
4                         & 2                            & 82.6                         & 82.6          & 90.8                         & 89.7          \\
8                         & 1                            & \textbf{84.0}                & \textbf{84.0} & 90.9                         & 89.9          \\ \hline
\end{tabular}
\end{table}
\subsection{Comparison with Other Methods}
The comparison results are shown in Tab.~\ref{SOTA}. 
Compared to the single-modal models, directly concatenating multi-modal features (row 3 in Tab.~\ref{SOTA}) improves $AP_{50}$ and $AP_{75}$ by more than 15\%. 
This proves that complementary information exists among different modalities.
For a fair comparison with other fusion methods, we embed their fusion modules into our framework so that different approaches can be validated in the same environment. 
CMML\cite{yang2019comprehensive} and CEN\cite{wang2022channel} merge the multi-modal features or pick one of them by automatically generating channel-wise weights for each modality.  
They score higher $AP$ in the validation set but lower one in the test set than "Concatenate". 
This may be because the generated weights are biased to make similar decisions to the source domain, thereby reducing model generalization in the external data. 
Moreover, CMF only highlights similar features between two modalities, ignoring that each modality contains some unique features. 
TMM focuses on both modality-specific and modality-invariant information, but the chaotic confusion of the two types of information deteriorates the model performance. 
Therefore, both CMF\cite{wang2022channel} and TMM\cite{li2022transiam} fail to outperform weight-based models. 
On the contrary, our AMF module prevents information entanglement by conducting cross-attention and self-attention blocks in parallel. 
It achieves $AP_{50}$=82.8, $AP_{75}$=60.6 in the validation set and $AP_{50}$=79.5, $AP_{75}$=39.2 in the test set, outperforming all competing methods while demonstrating superior generalization ability. 
Meanwhile, the improvement of detection performance is beneficial to our OTA module to obtain lesion features from more precise locations, thereby improving the accuracy of benign and malignant renal tumor diagnosis.

\begin{table}[]\centering
\caption{Diagnosis results of different methods.}\label{SOTA}
\begin{tabular}{|c|cccc|cccc|}
\hline
\multirow{2}{*}{Fusion Methods} & \multicolumn{4}{c|}{Validation}                                                                           & \multicolumn{4}{c|}{External Test}                                                                        \\ \cline{2-9} 
                                & \multicolumn{1}{c|}{$AP_{50}$} & \multicolumn{1}{c|}{$AP_{75}$}          & \multicolumn{1}{c|}{Accuracy} & F1-score & \multicolumn{1}{c|}{$AP_{50}$} & \multicolumn{1}{c|}{$AP_{75}$}          & \multicolumn{1}{c|}{Accuracy} & F1-score \\ \hline
B-mode                          & 59.6                      & \multicolumn{1}{c|}{38.0}          & 76.4                          & 76.3     & 58.1                      & \multicolumn{1}{c|}{22.1}          & 80.4                          & 79.1     \\ \cline{1-1}
CEUS-mode                       & 60.1                      & \multicolumn{1}{c|}{30.6}          & 78.2                          & 78.1     & 52.1                      & \multicolumn{1}{c|}{14.1}          & 70.5                          & 69.3     \\ \hline
Concatenate                     & 78.8                      & \multicolumn{1}{c|}{50.5}          & 79.6                          & 79.5     & 76.8                      & \multicolumn{1}{c|}{38.8}          & 86.8                          & 85.7     \\ \cline{1-1}
CMML\cite{yang2019comprehensive}                            & 80.7                      & \multicolumn{1}{c|}{54.4}          & 80.1                          & 80.1     & 76.0                      & \multicolumn{1}{c|}{37.2}          & 87.4                          & 86.2     \\ \cline{1-1}
CEN\cite{wang2022channel}                             & 81.4                      & \multicolumn{1}{c|}{56.2}          & 83.0                          & 83.0     & 74.3                      & \multicolumn{1}{c|}{36.3}          & 85.1                          & 83.8     \\ \cline{1-1}
CMF\cite{xu2022remixformer}                             & 81.4                      & \multicolumn{1}{c|}{54.8}          & 79.7                          & 79.7     & 75.2                      & \multicolumn{1}{c|}{35.2}          & 87.8                          & 86.8     \\ \cline{1-1}
TMM\cite{li2022transiam}                             & 80.8                      & \multicolumn{1}{c|}{52.7}          & 80.1                          & 80.1     & 74.3                      & \multicolumn{1}{c|}{37.0}          & 84.4                          & 83.2     \\ \cline{1-1}
\textbf{Ours}                   & \textbf{82.8}             & \multicolumn{1}{c|}{\textbf{60.6}} & \textbf{84.0}                          & \textbf{84.0}     & \textbf{79.5}             & \multicolumn{1}{c|}{\textbf{39.2}} & \textbf{90.9}                          & \textbf{89.9}     \\ \hline
\end{tabular}
\end{table}

\section{Conclusions} 
In this paper, we create the first multi-modal CEUS video dateset and propose a novel attention-based multi-modal video fusion framework for renal tumor diagnosis using B-mode and CEUS-mode US videos. It encourages interactions between different modalities via a weight-sharing dual-branch backbone and automatically captures the modality-invariant and modality-specific information by the AMF module. It also utilizes a portable OTA module to aggregate information in the temporal dimension of videos, making video-level decisions. The design of the AMF module and OTA module is plug-and-play and could be applied to other multi-modal video tasks. The experimental results show that the proposed method outperforms single-modal, single-frame, and other state-of-the-art multi-modal approaches. 
\subsubsection{Data use declaration and acknowledgment.} Our dataset was collected from The Affiliated Nanchong Central Hospital of North Sichuan Medical College, Shenzhen People's Hospital, and Fujian Provincial Hospital hospitals. This study was approved by local institutional review boards.

%
%
%
%
\bibliographystyle{splncs04}
\bibliography{reference}

\begin{thebibliography}{10}
\providecommand{\url}[1]{\texttt{#1}}
\providecommand{\urlprefix}{URL }
\providecommand{\doi}[1]{https://doi.org/#1}

\bibitem{azam2022review}
Azam, M.A., Khan, K.B., Salahuddin, S., Rehman, E., Khan, S.A., Khan, M.A.,
  Kadry, S., Gandomi, A.H.: A review on multimodal medical image fusion:
  Compendious analysis of medical modalities, multimodal databases, fusion
  techniques and quality metrics. Computers in biology and medicine
  \textbf{144},  105253 (2022)

\bibitem{chen2021domain}
Chen, C., Wang, Y., Niu, J., Liu, X., Li, Q., Gong, X.: Domain knowledge
  powered deep learning for breast cancer diagnosis based on contrast-enhanced
  ultrasound videos. IEEE Transactions on Medical Imaging  \textbf{40}(9),
  2439--2451 (2021)

\bibitem{chen2019multi}
Chen, H., Li, Y., Su, D.: Multi-modal fusion network with multi-scale
  multi-path and cross-modal interactions for rgb-d salient object detection.
  Pattern Recognition  \textbf{86},  376--385 (2019)

\bibitem{fang2022weighted}
Fang, J., Li, A., OuYang, P.Y., Li, J., Wang, J., Liu, H., Xie, F.Y., Liu, J.:
  Weighted concordance index loss-based multimodal survival modeling for
  radiation encephalopathy assessment in nasopharyngeal carcinoma radiotherapy.
  In: Medical Image Computing and Computer Assisted Intervention--MICCAI 2022:
  25th International Conference, Singapore, September 18--22, 2022,
  Proceedings, Part VII. pp. 191--201. Springer (2022)

\bibitem{ge2021yolox}
Ge, Z., Liu, S., Wang, F., Li, Z., Sun, J.: Yolox: Exceeding yolo series in
  2021. arXiv preprint arXiv:2107.08430  (2021)

\bibitem{george2022analysis}
George, M., Anita, H.: Analysis of kidney ultrasound images using deep learning
  and machine learning techniques: A review. Pervasive Computing and Social
  Networking: Proceedings of ICPCSN 2021 pp. 183--199 (2022)

\bibitem{huang2022personalized}
Huang, H., Dong, Y., Jia, X., Zhou, J., Ni, D., Cheng, J., Huang, R.:
  Personalized diagnostic tool for thyroid cancer classification using
  multi-view ultrasound. In: Medical Image Computing and Computer Assisted
  Intervention--MICCAI 2022: 25th International Conference, Singapore,
  September 18--22, 2022, Proceedings, Part III. pp. 665--674. Springer (2022)

\bibitem{kapetas2019quantitative}
Kapetas, P., Clauser, P., Woitek, R., Wengert, G.J., Lazar, M., Pinker, K.,
  Helbich, T.H., Baltzer, P.A.: Quantitative multiparametric breast ultrasound:
  application of contrast-enhanced ultrasound and elastography leads to an
  improved differentiation of benign and malignant lesions. Investigative
  radiology  \textbf{54}(5), ~257 (2019)

\bibitem{li2022transiam}
Li, X., Ma, S., Tang, J., Guo, F.: Transiam: Fusing multimodal visual features
  using transformer for medical image segmentation. arXiv preprint
  arXiv:2204.12185  (2022)

\bibitem{lin2022new}
Lin, Z., Lin, J., Zhu, L., Fu, H., Qin, J., Wang, L.: A new dataset and a
  baseline model for breast lesion detection in ultrasound videos. In: Medical
  Image Computing and Computer Assisted Intervention--MICCAI 2022: 25th
  International Conference, Singapore, September 18--22, 2022, Proceedings,
  Part III. pp. 614--623. Springer (2022)

\bibitem{liu2017medical}
Liu, Y., Chen, X., Cheng, J., Peng, H.: A medical image fusion method based on
  convolutional neural networks. In: 2017 20th international conference on
  information fusion (Fusion). pp.~1--7. IEEE (2017)

\bibitem{ljungberg2019european}
Ljungberg, B., Albiges, L., Abu-Ghanem, Y., Bensalah, K., Dabestani, S.,
  Fern{\'a}ndez-Pello, S., Giles, R.H., Hofmann, F., Hora, M., Kuczyk, M.A.,
  et~al.: European association of urology guidelines on renal cell carcinoma:
  the 2019 update. European urology  \textbf{75}(5),  799--810 (2019)

\bibitem{shi2022yolov}
Shi, Y., Wang, N., Guo, X.: Yolov: Making still image object detectors great at
  video object detection. arXiv preprint arXiv:2208.09686  (2022)

\bibitem{wang2021transbts}
Wang, W., Chen, C., Ding, M., Yu, H., Zha, S., Li, J.: Transbts: Multimodal
  brain tumor segmentation using transformer. In: Medical Image Computing and
  Computer Assisted Intervention--MICCAI 2021: 24th International Conference,
  Strasbourg, France, September 27--October 1, 2021, Proceedings, Part I 24.
  pp. 109--119. Springer (2021)

\bibitem{wang2020deep}
Wang, Y., Huang, W., Sun, F., Xu, T., Rong, Y., Huang, J.: Deep multimodal
  fusion by channel exchanging. Advances in neural information processing
  systems  \textbf{33},  4835--4845 (2020)

\bibitem{wang2022channel}
Wang, Y., Sun, F., Huang, W., He, F., Tao, D.: Channel exchanging networks for
  multimodal and multitask dense image prediction. IEEE Transactions on Pattern
  Analysis and Machine Intelligence  (2022)

\bibitem{xu2022remixformer}
Xu, J., Gao, Y., Liu, W., Huang, K., Zhao, S., Lu, L., Wang, X., Hua, X.S.,
  Wang, Y., Chen, X.: Remixformer: A transformer model for precision skin tumor
  differential diagnosis via multi-modal imaging and non-imaging data. In:
  Medical Image Computing and Computer Assisted Intervention--MICCAI 2022: 25th
  International Conference, Singapore, September 18--22, 2022, Proceedings,
  Part III. pp. 624--633. Springer (2022)

\bibitem{yang2019comprehensive}
Yang, Y., Wang, K.T., Zhan, D.C., Xiong, H., Jiang, Y.: Comprehensive
  semi-supervised multi-modal learning. In: IJCAI. pp. 4092--4098 (2019)

\bibitem{zhu2022contrast}
Zhu, J., Li, N., Zhao, P., Wang, Y., Song, Q., Song, L., Li, Q., Luo, Y.:
  Contrast-enhanced ultrasound (ceus) of benign and malignant renal tumors:
  Distinguishing ceus features differ with tumor size. Cancer Medicine  (2022)

\end{thebibliography}

\end{document}